\documentclass[twocolumn,floatfix,prb,aps,showpacs,bibliography]{revtex4-1}
\usepackage{graphicx,amsmath,amssymb,color}
\usepackage{nicefrac}
\usepackage{url}
\usepackage[titletoc,title]{appendix}

\newcommand{\be}{\begin{equation}}
\newcommand{\ee}{\end{equation}}

\newcommand{\ba}{\begin{eqnarray}}
\newcommand{\ea}{\end{eqnarray}}

\renewcommand{\vec}[1]{{\textbf{\textit{#1}}}}

\begin{document}

\title{Exact results for model wave functions of anisotropic composite fermions in the fractional quantum Hall effect}
\author{Ajit C. Balram and J. K. Jain}

\affiliation{Department of Physics, 104 Davey Lab, Pennsylvania State University, University Park, Pennsylvania 16802, USA}
  
\begin{abstract} 
The microscopic wave functions of the composite fermion theory can incorporate electron mass anisotropy by a trivial rescaling of the coordinates. These wave functions are very likely adiabatically connected to the actual wave functions of the anisotropic fractional quantum Hall states. We show in this paper that they possess the nice property that their energies can be analytically related to the previously calculated energies for the isotropic states through a universal scale factor, thus allowing an estimation of several observables in the thermodynamic limit for all fractional quantum Hall states as well as the composite fermion Fermi sea. The rather weak dependence of the scale factor on the anisotropy provides insight into why fractional quantum Hall effect and composite fermions are quite robust to electron mass anisotropy. We discuss how better, though still approximate, wave functions can be obtained by introducing a variational parameter, following Haldane [F. D. M. Haldane, Phys. Rev. Lett. {\bf 107}, 116801 (2011)], but the resulting wave functions are not readily amenable to calculations. Our considerations are also applicable, with minimal modification, to the case where the dielectric function of the background material is anisotropic.
\end{abstract}
\maketitle

\section{Introduction}
Striking recent experiments from Princeton University by Gokmen {\em et al.}\cite{Gokmen10}, Kamburov {\em et al.}\cite{Kamburov13,Kamburov14}, and Mueed {\em et al.}\cite{Mueed15b,Mueed15c} have investigated the role of electron mass anisotropy on the composite fermion (CF) Fermi sea, which was predicted in the early 1990s by Halperin, Lee, and Read\cite{Halperin93} and Kalmeyer and Zhang\cite{Kalmeyer92} and has been studied extensively~\cite{Willett93,Kang93,Goldman94,Smet96,Smet98,Willett99,Smet99,Freytag02,Kamburov12}. Anisotropy in the mass for electrons (at zero magnetic field) can occur due to the band structure, e.g. in AlAs quantum wells\cite{Shayegan06} where the Fermi pockets are elliptical with the longitudinal and transverse electron band masses that differ by a factor of $\sim$5, or can be produced by application of an in-plane magnetic field in finite width GaAs quantum wells\cite{Kamburov12b,Kamburov13b,Mueed15a,Mueed15c}. Direct measurements of the CF Fermi wave vectors in perpendicular directions\cite{Kamburov13,Kamburov14}  demonstrate that the CF Fermi sea also becomes anisotropic. These experiments also measure the degree of anisotropy for the CF Fermi sea as a function of the anisotropy of the electron Fermi sea.  In isotropic systems, the electron band mass $m_b$ is not a parameter for the CF physics, to the extent one can neglect Landau-level (LL) mixing, and the effective mass of composite fermions is determined entirely by the interelectron interactions\cite{Halperin93,Murthy03,Jain07}. The experiments thus raise many fundamental questions: How are composite fermions and their compressible and incompressible states modified by a mass anisotropy of the underlying electrons? How can one incorporate the effect of mass anisotropy into the CF theory? How robust are composite fermions to electron band mass anisotropy? What is the physics of the state obtained at very large anisotropies when composite fermions are destabilized?

Balagurov and Lozovik\cite{Balagurov00} had already addressed this issue theoretically many years ago for the CF Fermi sea using the Chern-Simons theory\cite{Lopez91,Halperin93}. Significant progress has been made after a recent work by Haldane\cite{Haldane11}, and many recent papers\cite{Qiu12,Qiu13,Yang12b,Yang12c,Yang13,Papic13,Murthy13,Apalkov14,Ghazaryan15,Johri15} have investigated fractional quantum Hall effect (FQHE) states and the Fermi sea in systems with anisotropy. Somewhat conflicting conclusions have been reached for the CF Fermi sea: Ref.~\cite{Balagurov00} finds that the CF Fermi contour is identical to the electron Fermi contour at zero magnetic field, whereas Yang~\cite{Yang13} and Murthy~\cite{Murthy13} find a smaller anisotropy for the former. 

Most of the studies in this context have focused on the Laughlin 1/3 state\cite{Laughlin83}, and numerical work has been restricted to fairly small systems. We ask in this work how the CF theory can be modified to include electron mass anisotropy. That would shed light on a broad class of FQHE states, their excitations, the CF Fermi sea\cite{Halperin93}, and many other phenomena, and will tell us how the vortex structure of composite fermions is altered in the presence of a mass anisotropy. We present in this work a step in that direction. We begin by noting that the wave functions of composite fermions\cite{Jain89} can be readily generalized to anisotropic systems by a redefinition of the coordinates. These wave functions are expected to be adiabatically connected to the actual ground state so long as the system is in the gapped FQHE phase. We show that for these wave functions, the energies for many quantities of interest are modified by an anisotropy-dependent scale factor, which allows one to export many previously known results to anisotropic systems. These include ground state energies, transport gaps, spin polarization phase diagram, etc. To further improve upon these wave functions, we show how a variational parameter can be introduced for composite fermions, following Haldane\cite{Haldane11}. The form of the resulting wave functions, however, is complex and not amenable to immediate calculations. (This is the case even for the 1/3 state for which only very small systems have been studied in the literature.\cite{Yang12b,Yang12c}) We note that our results are also applicable to the case where the interaction is anisotropic (as would be induced from an anisotropic dielectric function of the host material).

We find it convenient to formulate the generalization of the CF theory in the disk geometry. The spherical geometry\cite{Haldane83}, in which electrons move on the surface of a sphere subjected to a radial magnetic field, is not appropriate for the issue of anisotropic mass. The torus geometry\cite{Haldane85,Rezayi00} appears most appropriate for this question (and is indeed very convenient for numerical studies of the anisotropy\cite{Wang12,Yang12c}), but the construction of Jain's CF wave functions in this geometry is nontrivial even for a system with isotropic electron mass\cite{Hermanns08,Hermanns13} (because the product of two LL wave functions is not a valid wave function). 
We note that there has also been much experimental and theoretical work on anisotropic stripe, liquid crystal, and FQHE states (see, for example, Refs.\cite{Lilly99,Xia11,Liu13,Fogler96,Fradkin99,Mulligan10,Mulligan11,Shi15,Samkharadze15}) for systems with isotropic electron mass.

The plan of the paper is as follows. In Sec. \ref{acf}, we show that the wave functions for various FQHE states can be modified straightforwardly to accommodate electron mass anisotropy. We show that for these wave functions, many experimentally measurable quantities for anisotropic systems are related to those for isotropic systems by a simple scale factor. In Sec. \ref{variational} we introduce a variational parameter, following Haldane, which can produce a better quantitative approximation for the actual anisotropic state. We find, however, that these states are very complex and not amenable to computations. The paper is concluded in Sec. IV.

\section{Anisotropic composite fermions -- zeroth order description}
\label{acf}
To establish the notation, we begin with the single particle problem:
\be
H_0={1\over 2m_x}\left( p_x-{e\over c}A_x \right)^2+{1\over 2m_y}\left( p_y-{e\over c}A_y \right)^2
\ee
We use the symmetric gauage in which $\vec{A}=(B/2)(-y, x, 0)$ and define $l=\sqrt{\hbar c/eB}$, $l_x=\sqrt{\gamma}\,l$, $l_y=l/\sqrt{\gamma}$, $\omega_c=eB/[c(m_x m_y)^{1/2}]$, $x_s=x/l_x$, $y_s=y/l_y$, where the mass anisotropy parameter is
\be
\gamma=\sqrt{m_y/m_x}=l_x/l_y.
\ee
We further define dimensionless complex coordinates 
\be
z=x_{s}-iy_{s}= {x\over l_x}-i {y\over l_y},\;\; \overline{z}=x_s+iy_s={x\over l_x}+i {y\over l_y}
\label{zdef}
\ee
where the scaled distances are denoted by the subscript $s$,
and introduce the standard ladder operators 
\begin{eqnarray}
a&= (z/2+2\overline{\partial})/\sqrt{2}, \; a^\dagger= (\overline{z}/2-2{\partial})/\sqrt{2}\\
b&= (\overline{z}/2+2{\partial})/\sqrt{2},\;b^{\dagger}= (z/2-2\overline{\partial})/\sqrt{2}
\end{eqnarray}
where $\partial=\partial/\partial z$ and $\overline{\partial}=\partial/\partial \overline{z}$. These have 
commutators $[a,a^{\dagger}]=1$, $[b,b^{\dagger}]=1$, and all other commutators vanish.
The Hamiltonian can be recast (taking $\hbar \omega_c=1$) as 
$H_0=a^\dagger a+1/2$, 
which has eigenstates given by $|n,m\rangle$ with eigenvalues $(n+1/2)$ and wave functions
\be
\phi_{n,m}= {1\over \sqrt{2 \pi}} {   (b^{\dagger})^m \over \sqrt{m!}  } {   (a^{\dagger})^n \over \sqrt{n!}  } e^{-{z\overline{z}\over4}}
\ee
where $n=0, 1, 2, \cdots$ and $m=0, 1, 2, \cdots$. 
Even though $m$ loses its interpretation as the $z$ component of the angular momentum, it can still be used as a label of the single-particle orbitals. In particular, the lowest LL (LLL) wave functions are given by $z^m e^{-\overline{z}z/4l^2}$, with maxima along an elliptical contour enclosing an area $2m\pi l^2$, with radii along the $x$ and $y$ axes given by $\sqrt{2m} l_x$ and $\sqrt{2m}l_y$. A note regarding notation ought to be stressed: $z$ and $\bar{z}$ are defined in terms of the scaled variables in what follows below. 

We now turn to FQHE, where electrons interact via the \emph{isotropic} Coulomb interaction $e^2/\epsilon r$. The wave functions of FQHE states at the fractions of the form $\nu=n/(2pn\pm 1)$ for an isotropic system are given by\cite{Jain89}: 
\be
\Psi_{n\over 2pn+1}={\cal P}_{\rm LLL} \Phi_n(z,\overline{z}) \prod_{j<k}(z_j-z_k)^{2p}
\label{Psi01}
\ee
\be
\Psi_{n\over 2pn-1}={\cal P}_{\rm LLL} [\Phi_n(z,\overline{z})]^* \prod_{j<k}(z_j-z_k)^{2p}
\label{Psi02}
\ee
where $\Phi_n$ is the wave function of $n$ filled Landau levels and the operator ${\cal P}_{\rm LLL}$ implements LLL projection, which can be accomplished as explained in Refs.~\cite{Dev92a,Jain97,Jain97b,Moller05,Jain07,Davenport12,Mukherjee15b}. Wave functions can be constructed analogously for the charged and neutral excitations of the FQHE states\cite{Dev92,Scarola00}, and for FQHE states involving spin and / or valley degree of freedom\cite{Wu93,Park98,Jain07}. 

It is evident that, with $z$ defined as in Eq.~\ref{zdef}, these are all valid LLL wave functions even in the presence of mass anisotropy and we call these ``zeroth-order'' wave functions. The integer quantum Hall effect (IQHE) states, the LLL projection, etc. all go through as for the isotropic case. They also satisfy some of the properties that originally motivated these wave functions in the absence of anisotropy\cite{Jain89}. Briefly, these establish a mapping between the FQHE and the IQHE. Experiments show that such a mapping must exist even in the presence of (at least sufficiently weak) anisotropy, given that the prominent sequences $n/(2n\pm 1)$ and the $1/2$ Fermi sea are observed even in the presence of anisotropy. If we neglect the LLL projection, these wave functions explicitly build very good correlations between electrons by ensuring that the probability of electrons approaching one another vanishes very rapidly as $r^{4p+2}$, where $r$ is the distance between them. Analogous to the isotropic case, the amplitude outside of the LLL is small even without LLL projection\cite{Jain89,Trivedi91,Kamilla97b}, and therefore one may expect that the LLL projection does not destroy these correlations. 

Because these wave functions are adiabatically connected to the exact FQHE states, it is worth asking what they imply for various measurable quantities. We show the useful result that, for the $1/r$ interaction (which is relevant for electrons in the LLL), many observable quantities of the wave functions in Eqs. \ref{Psi01} and \ref{Psi02} are related to those of the isotropic systems through an overall scale factor, which allows one to borrow known results from the extensive literature for the CF states in isotropic systems. We believe that our results give insight into the qualitative effect of electron band mass anisotropy, and also provide a first approximation for the quantitative corrections to various quantities. 

\subsection{CF Fermi sea} We note, to begin with, that the above generalization immediately allows us to make a statement regarding the anisotropy of the CF Fermi sea\cite{Halperin93}.  The pair-correlation function of the isotropic CF Fermi sea exhibits $\sin(2 k_F r+\phi)$ Friedel oscillations~\cite{Kamilla97}, where $k_F \approx \sqrt{4\pi \rho}$, as appropriate for a fully spin polarized system\cite{Balram15b}. For $\nu=1/2$ CF Fermi sea, with $\rho=\nu /2\pi l^2$, the oscillations are given by $\sin(2 r/l+\phi)$. This result carries over to the anisotropic case in scaled units. Converting to laboratory units, the oscillations behave as $\sin(2 x_{s}+\phi)=\sin(2 k^*_{x,F} x+\phi)$ along the $x$ direction and as $\sin(2 y_{s}+\phi)=\sin(2 k^*_{y,F} y+\phi)$ for the $y$ direction where 
\be k^*_{x,F}={1\over l_x}={k_F \over \sqrt{\gamma}},\;k^*_{y,F}={1\over l_y}=\sqrt{\gamma} k_F.
\ee The ratio of the CF Fermi wave vectors is given by 
\be
{k^*_{x,F}   \over  k^*_{y,F} }={1 \over \gamma}=\left( m_x \over m_y \right)^{1/2}
\label{anisotropy}
\ee
This is the same ratio as that for electrons in zero magnetic field, indicating that the CF Fermi sea has the same contour as the electron Fermi sea at zero magnetic field, in agreement with Balagurov and Lozovik\cite{Balagurov00}.

This conclusion appears, at first, to be inconsistent with the experiments of Kamburov {\em et al.}\cite{Kamburov14}, who find a smaller anisotropy for the CF Fermi sea than for the electron Fermi sea. In fact, they find that even when the electron Fermi sea becomes peanut shaped or breaks into two pieces (i.e., becomes bilayer like), the CF Fermi sea remains elliptical, with much smaller anisotropy. However, a closer look shows that not necessarily to be the case.
Kamburov {\em et al.}~\cite{Kamburov14} have noted that the physics of electrons at zero magnetic field and of composite fermions at $\nu=1/2$ differ in an important aspect. The electron band mass is very nearly the same in the directions parallel and perpendicular to the two-dimensional layer, whereas for composite fermions the two masses are vastly different: the CF mass parallel to the plane is a result of interelectron interactions, and has been determined\cite{Du93,Manoharan94,Kamburov14} to be approximately equal to the electron mass in vacuum ($m_e$), whereas the physics in the transverse direction is determined by the electron band mass $m_b$. The Fermi sea anisotropy induced by application of a parallel magnetic field depends sensitively on the ratio of the perpendicular and parallel masses.
Kamburov {\em et al.}\cite{Kamburov14} have considered a system of auxiliary fermions at zero magnetic field that have different $m_\parallel$ and $m_\perp$, and obtained $k_{x,F}   \over  k_{y,F}$ as a function of an in-plane field $B_x$ in a perturbative calculation to be $(1-\zeta B_x^2m_z/m_\parallel)^{1/2}$ where $\zeta$ depends on the quantum-well width. They find this to be in good quantitative agreement with the observed $k^*_{x,F}   \over  k^*_{y,F}$ for composite fermions if they take $m_\parallel=m_e$ and $m_\perp=m_b=0.067 m_e$.  The observations of Kamburov {\em et al.}\cite{Kamburov14} are thus not inconsistent with Eq.~\ref{anisotropy}, provided one accounts for different $m_\parallel$ and $m_\perp$. An experimental measurement of CF Fermi contour in a material such as an AlAs quantum well, where the anisotropy of the electron band mass does not derive from a parallel field, would provide a more direct test of the relation between the anisotropies of the electron and the CF Fermi contours.  

\subsection{Ground state energies} 
For electrons in the presence of a background charge, the total energy is the sum of the electron-electron, electron-background, and background-background interaction terms\cite{Giuliani08}:
\be
\langle \hat{V}_{\rm ee} \rangle={e^2 \over 2} \int d\vec{r}_{1} \int d\vec{r}_{2}~\rho(\vec{r}_{1}) \rho(\vec{r}_{2}) g(\vec{r}_{1},\vec{r}_{2}) V(\vec{r}_{1},\vec{r}_{2})
\label{ee}
\ee
\be
\langle \hat{V}_{\rm eb} \rangle=-{e^2} \int d\vec{r}_{1} \int d\vec{r}_{2}~\rho(\vec{r}_{1}) \rho_b(\vec{r}_{2}) V(\vec{r}_{1},\vec{r}_{2})
\label{eb}
\ee
\be
\langle \hat{V}_{\rm bb} \rangle={e^2 \over 2} \int d\vec{r}_{1} \int d\vec{r}_{2}~\rho_b(\vec{r}_{1}) \rho_b(\vec{r}_{2}) V(\vec{r}_{1},\vec{r}_{2})
\label{bb}
\ee
where $-e\rho(\vec{r})$ and $e\rho_b(\vec{r})$ are the electron and background charge densities and $g(\vec{r}_{1},\vec{r}_{2})$ is the pair correlation function defined as:
\be
g(\vec{r}_{1},\vec{r}_{2})=\frac
{\langle \hat{\Psi}^{\dagger}(\vec{r}_{1})  \hat{\Psi}^{\dagger}(\vec{r}_{2})  \hat{\Psi}(\vec{r}_{2}) \hat{\Psi}(\vec{r}_{1})    \rangle}{\rho(\vec{r}_{1}) \rho(\vec{r}_{2})}
\ee
where $\hat{\Psi}^{\dagger}(\vec{r})$ is the electron field creation operator.

We next transform to the scaled variables $\vec{r}_{s}=(x_{s},y_{s})$. In the scaled frame, the incompressible ground state wave function has both translational and rotational invariance and the pair correlation function depends only on $|\vec{r}_{1s}-\vec{r}_{2s}|=r_{s}$. (In the scaled variables, the mass anisotropy reflects through a direction dependent interaction.) The total energy obtained by adding the above three terms assuming uniform density $\rho(\vec{r})=\rho_b(\vec{r})=\rho_0$ as is relevant for an infinite system simplifies to
\begin{eqnarray}
\langle \hat{V} \rangle&=&\langle \hat{V}_{\rm ee} \rangle+\langle \hat{V}_{\rm eb} \rangle+\langle \hat{V}_{\rm bb} \rangle \\
&=&{\rho_0^2 e^2\over 2}  \int d\vec{r}_{1s} \int d\vec{r}_{2s}~[g(r_{s})-1] V(\vec{r}_{1s},\vec{r}_{2s}) \nonumber 
\end{eqnarray}

We will now specialize to the Coulomb interaction $1/\epsilon|\vec{r}_1-\vec{r}_2|$, which in the scaled coordinates is given by:
\be
V(\vec{r}_{1s},\vec{r}_{2s})=\frac{1}{\epsilon\sqrt{(\sqrt{\gamma}r_{s}\cos\theta)^2+(\frac{r_{s}\sin\theta}{\sqrt{\gamma}})^2}} 
\ee
where $r_{s}$ and $\theta$ are the polar coordinates of $\vec{r}_{1s}-\vec{r}_{2s}$.

\begin{figure}
\begin{center}
\includegraphics[width=0.45\textwidth]{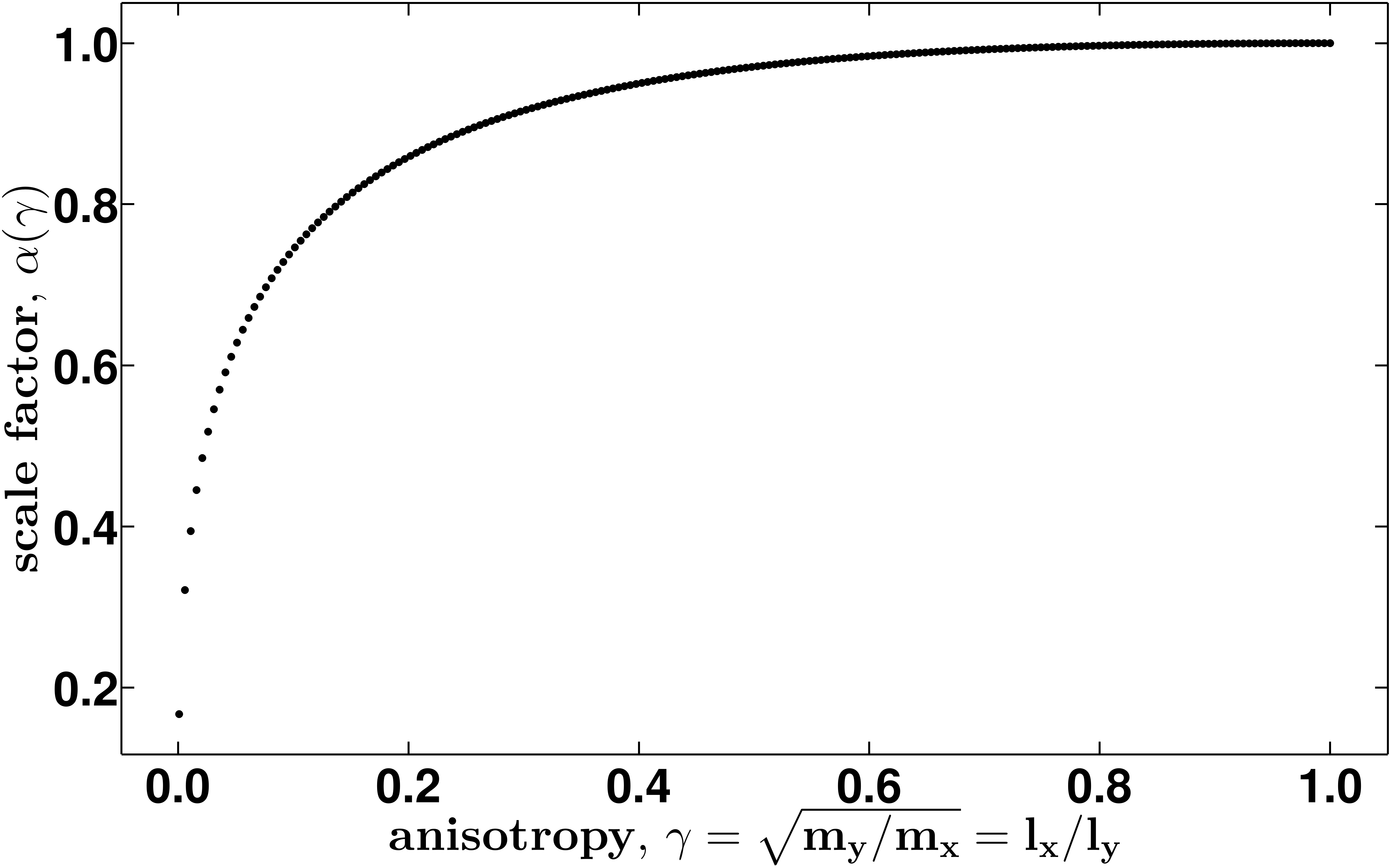}
\caption{The scale factor $\alpha$ as a function of the anisotropy parameter $\gamma=l_{x}/l_{y}=\sqrt{m_{y}/m_{x}}$. We only show the region $0<\gamma\leq 1$ since $\alpha(\gamma)=\alpha(1/\gamma)$.}
\label{fig:1}
\end{center}
\end{figure}

By switching to center of mass and relative coordinates in the scaled variables we obtain:
\begin{eqnarray}
\frac{\langle \hat{V} \rangle}{N}=\frac{e^2\rho_{0}}{2\epsilon}\int d\vec{r}_{s}  \frac{ [g(r_{s})-1]}{\sqrt{(\sqrt{\gamma}r_{s}\cos\theta)^2+(\frac{r_{s}\sin\theta}{\sqrt{\gamma}})^2}} 
\end{eqnarray}
Performing the angular integrals, this reduces to 
\be
{{\langle \hat{V} \rangle}\over N} = \alpha(\gamma) \left[{{\langle \hat{V} \rangle}\over N}\right]_{\gamma=1}
\label{energy}
\ee
with the scale factor
\be
\alpha(\gamma)=\frac{1}{\pi\sqrt{\gamma}} \Bigg[K\Big(1-\frac{1}{\gamma^2}\Big)+ \gamma K\Big(1-\gamma^2\Big) \Bigg]
\label{gamma}
\ee
where $K(m)$ is the complete elliptic integral of the first kind. The scale factor $\alpha(\gamma)$ is plotted in Fig.~\ref{fig:1}. We note that the scale factor $\alpha$ changes very little from unity even when the electron mass anisotropy $m_y/m_x$ becomes very large, which gives insight into the robustness of the various states of composite fermions to the electron mass anisotropy. 

In the above analysis we have assumed an infinite system with the wave function satisfying translational and rotational invariance in the scaled variables. In the Appendix \ref{appendix1} we show that the scaling of Eq.~\ref{energy} holds even for a finite droplet so long as the system has rotation symmetry in the scaled variables.

\subsection{Excitations gaps}

The excitation gaps measured in transport experiments are interpreted as the energy required to create a far separated pair of particle hole excitation of composite fermions. For the state at $\nu=n/(2pn+1)$ [$\nu=n/(2pn-1)$], the particle [hole] excitation consists of adding a composite fermion to the empty $(n+1)^{\rm th}$ CF Landau-like level (termed $\Lambda$L), and the hole [particle] excitation consists of removing a CF from the topmost filled $n^{\rm th}$ $\Lambda$L. In the scaled frame, these excitations have the same density profile as those for the isotropic system. In the laboratory frame, they are elongated in one direction while shrunk in the other. One can evaluate the energy gap by considering states containing a single particle, a single hole, and the ground state:
\be
\Delta_{\frac{n}{2pn\pm1}}(\gamma)=E^{qh}_{\frac{n}{2pn\pm1}}+E^{qp}_{\frac{n}{2pn\pm1}}-2E^{gs}_{\frac{n}{2pn\pm1}}
\label{largek}
\ee
Because the states do not have uniform density, we must work with the general expressions in Eqs. \ref{ee}, \ref{eb} and \ref{bb} for the evaluation of the terms on the right hand side. 

While the wave functions for excitations do not satisfy translational invariance, they do satisfy rotational invariance in the scaled frame of reference, for states in which the additional composite fermion in the $(n+1)^{\rm th}$ $\Lambda$L is taken to be located at the origin (with no loss of generality), or, similarly, the missing CF in the $n^{\rm th}$ $\Lambda$L is taken to be located at the origin. This leads to a simplification in our calculations. We show in the Appendix \ref{appendix1} that the energy of any wave function that satisfies rotational invariance in the scaled frame is related to the energy of the corresponding isotropic state by the same scale factor 
\be
\langle \hat{V} \rangle_{\gamma}=\alpha(\gamma)\langle \hat{V} \rangle_{\gamma=1}
\label{tot_rot_inv}
\ee
Hence we have
\be
\Delta_{\frac{n}{2pn\pm1}}(\gamma)=\alpha(\gamma)\Delta_{\frac{n}{2pn\pm1}}(\gamma=1)
\ee
and therefore gaps decrease with increasing anisotropy.

Very recent experiments by Mueed {\em et al.}\cite{Mueed15c} have reported a study of the gap of the fully spin polarized 2/3 fractional Hall state as a function of the anisotropy parameter. They find that for up to $\gamma=0.80$ there is essentially no change in the gap, as expected from theory. However, for $\gamma=0.66$ the gap is reduced by 40\% from its isotropic value. This is unexpected from our theory or from previous calculations\cite{Wang12}, and might be indicative of proximity to an instability (e.g. into a bilayer 2/3 phase).  

\subsection{CF mass}
The ``activation" mass $m_{\rm a}^*$ of composite fermions has been defined by interpreting the activation gap as an effective cyclotron energy $\hbar eB^*/m_{\rm a}^* c$, of composite fermions in an effective magnetic field $B^*$\cite{Halperin93,Du93,Manoharan94,Du94,Leadley94}. Since the gaps are reduced by anisotropy, the activation mass of composite fermions increases with increasing anisotropy as
\be
m_{\rm a}^*(\gamma)=[\alpha(\gamma)]^{-1}m_{\rm a}^*(\gamma=1)
\ee 
Note that the CF activation mass has no directional dependence. 

\subsection{Neutral collective mode}
The neutral excitations are excitons of composite fermions\cite{Dev92,Scarola00,Murthy99,Kukushkin09}. The wave vector of the excitation is proportional to the distance between the excited composite fermion and the hole left behind.  For large wave vectors the energy of the neutral modes is not expected to depend on the relative orientation of the constituent CF particle and CF hole and is simply given by Eq.~\ref{largek}, and thus depends only on the magnitude of $k$. In this limit, the energy of the neutral mode also scales by the factor $\alpha(\gamma)$ in the presence of anisotropy. For small wave vectors, when the particle and the hole excitations overlap, the energy of the collective mode is expected to depend on the direction. In particular, the energies at the roton minima\cite{Girvin85} will have directional dependence. We believe that some features of the small wave vector dispersion can be captured by scaling the wave-vector appropriately, but we have not calculated the dispersion explicitly. Apalkov and Chakraborty\cite{Apalkov14} have calculated the collective excitations of the Laughlin state in the $n=0$ and $1$ Landau levels of graphene in the presence of anisotropic interactions, and the calculated dispersion indeed shows directional dependence at small wave vectors, but none at large wave vectors. Yang \emph{et. al} have also calculated the dispersion of the collective mode in the presence of the effective mass anisotropy for the 1/3 state\cite{Yang12c}.

\subsection{Spin phase transition} We next ask how the spin phase diagram of composite fermions is affected by anisotropy. Transitions between differently spin polarized CF states can be caused by changing the Zeeman energy, and the critical Zeeman energies have been determined for many fractions for an isotropic system\cite{Eisenstein89,Du95,Yeh99,Kukushkin99,Kukushkin00,Melinte00,Tiemann12,Feldman12,Feldman13,Liu14,Liu14a}. (These results also apply to systems with valley splitting\cite{Bishop07,Padmanabhan09}.) These are generally consistent with theoretical results\cite{Wu93,Park98,Park98b,Park99,Park01,Davenport12,Balram15,Balram15a}.
The above considerations lead to the prediction that the critical Zeeman energies for a system with anisotropic electron mass are given by
\be
E_{\rm Z}^{\rm critical}(\gamma)=\alpha(\gamma) \;E_{\rm Z}^{\rm critical}(\gamma=1)
\ee
The study of the critical Zeeman energy as a function of anisotropy can in principle allow a direct test of the validity of our approach and, in particular, of how far the wave functions in Eqs.~\ref{Psi01} and \ref{Psi02} remain valid.

\section{Variational parameter}
\label{variational}

The wave functions in Eqs.~\ref{Psi01} and \ref{Psi02} do not contain any adjustable parameter. They can be improved by introducing a variational degree of freedom $\eta$, following Haldane\cite{Haldane11}. The essential idea is to introduce the ladder operators
\be
b_\eta={1\over \sqrt{1-|\eta|^2}} (b + \eta b^{\dagger}),\;
b_\eta^{\dagger}={1\over \sqrt{1-|\eta|^2}} (\eta^* b + b^{\dagger})
\ee
which also satisfy the commutator $[b_\eta, b_\eta^{\dagger}]=1$. These create single-particle eigenorbitals that are distorted relative to those created by the $b^\dagger$'s. The single-particle orbitals created by $b_\eta^{\dagger}$ are given by: 
\be
\phi_{n,m}^{\eta}= {1\over \sqrt{2 \pi}} {   (b_\eta^{\dagger})^m \over \sqrt{m!}  } {   (a^{\dagger})^n \over \sqrt{n!}  } 
(1-|\eta|^2)^{1 \over 4} e^{-{1\over 4}\overline{z}z-{\eta \over 4}z^2}
\ee
where $e^{-{1\over 4}\overline{z}z-{\eta \over 4}z^2}$ is annihilated by $a$ and $b_{\eta}$.

The next question is, how do we generalize the wave functions in Eqs. \ref{Psi01} and \ref{Psi02}? We first note that for the IQHE states, the wave function is given by
\be
\Phi_n^{\eta}=(1-|\eta|^2)^{N \over 4}e^{-{\eta \over 4}\sum_iz_i^2} \Phi_n
\ee
where $\Phi_n$ is the state in the absence of anisotropy and $N$ is the number of particles. This relies on the property of $ \Phi_n$ that if any orbital $\phi_{q,m}$ is occupied, then all orbitals $\phi_{q',m}$ with $q'< q$ are also occupied.

Let us now write the CF wave function in Eqs. \ref{Psi01} and \ref{Psi02} in a slightly different form, which is amenable to generalization. Noting that $z=2(a+b^{\dagger})$, we can write the composite fermionization factor as
\be
\prod_{j<k}(z_j-z_k)^{2p}\propto \prod_{j<k}\left(a_j+b_j^{\dagger}-a_k-b_k^{\dagger} \right)^{2p}
\ee
Eqs. \ref{Psi01} and \ref{Psi02} can be recast as
\be
\Psi_{n\over 2pn+1}={\cal P}_{\rm LLL} \prod_{j<k}\left(a_j+b_j^{\dagger}-a_k-b_k^{\dagger} \right)^{2p} \Phi_n(z,\overline{z}) 
\ee
\be
\Psi_{n\over 2pn-1}={\cal P}_{\rm LLL} \prod_{j<k}\left(a_j+b_j^{\dagger}-a_k-b_k^{\dagger} \right)^{2p} [\Phi_n(z,\overline{z})]^* 
\ee
For $n=1$, i.e. $\nu=1/(2p+1)$, the expression simplifies to
\be
\Psi_{1\over 2p+1}= \prod_{j<k}\left(b_j^{\dagger}-b_k^{\dagger} \right)^{2p} \Phi_1(z,\overline{z}) = \prod_{j<k}\left(  b^\dagger_j-b^\dagger_k \right)^{2p+1} e^{-{1\over 4}\sum_i z_i \overline{z}_i}
\ee

One can now introduce a variational parameter by making the replacement:
\be
b\rightarrow b_{\eta},\; b^{\dagger}\rightarrow b^{\dagger}_{\eta}, \; \Phi_n\rightarrow \Phi_n^\eta=(1-|\eta|^2)^{1 \over 4}e^{-{\eta \over 4}\sum_iz_i^2} \Phi_n
\ee
which gives:
\be
\Psi_{n\over 2pn+1}={\cal P}_{\rm LLL} \prod_{j<k}\left(a_j+b_{\eta,j}^{\dagger}-a_k-b_{\eta,k}^{\dagger} \right)^{2p} \Phi^\eta_n(z,\overline{z}) 
\ee
\be
\Psi_{n\over 2pn-1}={\cal P}_{\rm LLL} \prod_{j<k}\left(a_j+b_{\eta,j}^{\dagger}-a_k-b_{\eta,k}^{\dagger} \right)^{2p} [\Phi^\eta_n(z,\overline{z})]^* 
\ee
In particular, Laughlin's wave function\cite{Laughlin83} generalizes to 
\be
\Psi^\eta_{\frac{1}{m}}=\prod_{j<k}\left(  b_{\eta,j}^\dagger-b_{\eta,k}^\dagger \right)^m (1-|\eta|^2)^{1 \over 4}e^{-{1\over 4}\sum_i z_i \overline{z}_i-{\eta \over 4}\sum_iz_i^2}
\ee
which is identical to the form previously considered\cite{Yang12c}. In principle, the lowest Landau-level projection can be performed by the method\cite{Dev92a} of bringing all the factors of $\bar{z}$ to the left of the $z$'s in each term and replacing $\bar{z}$ by $2\partial/\partial z$ which acts on all terms except the Gaussian factor $e^{-{1\over 4}\sum_i z_i \overline{z}_i}$. An explicit evaluation of these wave functions has been possible only for very small system sizes. These wave functions are also not amenable to the Jain-Kamilla projection method\cite{Jain97,Jain97b}, which has proved very useful for isotropic FQHE states. 

\section{Discussion}
\label{discussion}
The wave functions in Eqs.~\ref{Psi01} and \ref{Psi02} are expected to be adiabatically connected to the actual wave functions so long as the system remains in an incompressible FQHE phase. For small anisotropies, one can obtain the dependence of energies on the anisotropy by writing $\gamma=1-\varepsilon$ ($\varepsilon\ll 1$) and Taylor expanding the interaction:
\begin{eqnarray}
V(x_{s},y_{s};\varepsilon)&=&\frac{1}{\epsilon\sqrt{\gamma x_{s}^2+\frac{y_{s}^2}{\gamma}}}=\frac{1}{\epsilon\sqrt{(1-\varepsilon) x_{s}^2+\frac{y_{s}^2}{(1-\varepsilon)}}} \nonumber \\
&=& \frac{1}{\epsilon\sqrt{x_{s}^2+y_{s}^2}}+\frac{x_{s}^2-y_{s}^2}{\epsilon(x_{s}^2+y_{s}^2)^{\frac{3}{2}}}\varepsilon
+\mathcal{O}(\varepsilon^2)
\end{eqnarray}
Given that the eigenfunctions $\{ | \Psi\rangle \}$ of the ``unperturbed" Hamiltonian (the first term on the right) are symmetric in the scaled coordinates $x_{s}$ and $y_{s}$, we have $\langle \Psi |x_{s}^2| \Psi \rangle = \langle\Psi |y_{s}^2|\Psi\rangle$, indicating that the correction to the energy is of order $\mathcal{O}(\varepsilon^2)$. This is consistent with the above since the Taylor expansion of the scale factor gives:
\begin{equation}
\alpha(\varepsilon)=1-\frac{\varepsilon^2}{16}+\mathcal{O}(\varepsilon^3) 
\end{equation}
Ref. \onlinecite{Qiu12} also argued that the Coulomb energies for the Laughlin state with and without the variational parameter differ only at second order in the anisotropy.

\begin{figure}[t]
\begin{center}
\includegraphics[width=0.45\textwidth]{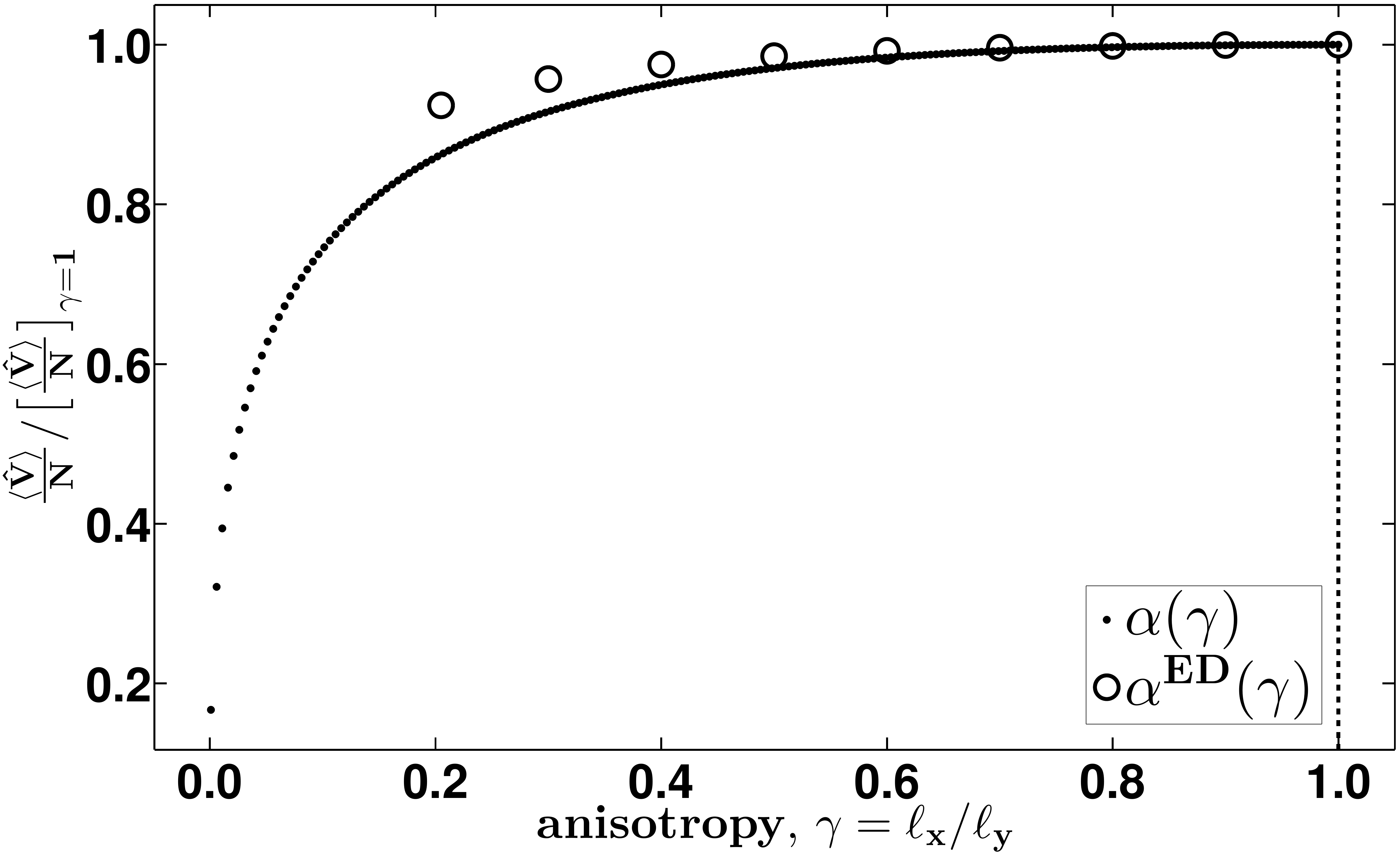}
\caption{A comparison of the ``exact'' scale factor $\alpha^{\rm ED}(\gamma)=\langle \hat{V} \rangle_{\gamma}/\langle \hat{V} \rangle_{\gamma=1}$  (open circles) with the scale factor $\alpha(\gamma)$ (filled dots) for the ``zeroth-order'' model wave functions as a function of the anisotropy parameter $\gamma=l_{x}/l_{y}$. The exact diagonalization data are reproduced from Fig. 1 of Yang {\em et al.}\cite{Yang12c}, and correspond to $N=7$ electrons at filling factor $\nu=1/3$ in the torus geometry.}
\label{fig:2}
\end{center}
\end{figure}

A detailed test of the accuracy of these wave functions will require comparing their energies to those obtained from exact diagonalization. In Fig. \ref{fig:2} we show the ``exact" scale factor $\alpha^{\rm ED}(\gamma)=\langle \hat{V} \rangle_{\gamma}/\langle \hat{V} \rangle_{\gamma=1}$ where the energies are obtained from exact diagonalization by Yang \emph{et. al.} \cite{Yang12c} in the torus geometry at $\nu=1/3$ for $N=7$ electrons. Although the exact $\alpha^{\rm ED}(\gamma)$ does not correspond to the thermodynamic limit, a comparison shows that our model wave functions capture the fact that the energy is very insensitive to effective mass anisotropy. We stress that our model wave functions do not capture the physics of the eventual instability of composite fermions with increasing anisotropy\cite{Wang12}. 

\section{Summary}
In summary, we have shown that wave functions for composite fermions can be modified straightforwardly to accommodate an anisotropic electron band mass. These ``zeroth-order" wave functions are adiabatically connected to the actual wave functions for the incompressible FQHE states, and thus provide a starting point for further investigation.  One of the nice features of these wave functions is that their Coulomb energies are related to those of the isotropic systems by an anisotropy-dependent scale factor, and consequently, many results known for the isotropic system can thus be transported straightforwardly to the system with anisotropy. The scale factor has very weak dependence on anisotropy for up to fairly large values of $m_x/m_y$, which gives insight into the robustness of compressible and incompressible states of composite fermions to electron mass anisotropy. It is possible to further improve upon these wave functions by introducing a variational parameter following Haldane\cite{Haldane11}, but the resulting wave functions are less amenable to explicit evaluations. Another possible way to introduce a variational parameter is following the work of Ref. \cite{Musaelian96}, which we have not pursued.

{\bf Acknolwedgment}
We acknowledge M. Shayegan for many helpful discussions. We thank E. H. Rezayi for sharing with us the data shown in Fig.~1 of Ref.~\onlinecite{Yang12c}. This work was supported by the U. S. NSF Grant No. DMR-1401636. We acknowledge the Research Computing and Cyberinfrastructure at Pennsylvania State University, which is in part funded by the NSF Grant No. OCI-0821527.

\begin{appendices}
\begin{widetext}
\section{Energy scaling of wave functions related to rotationally invariant states}
\label{appendix1}

In this appendix we show that the relation in Eq.~\ref{tot_rot_inv}:
\begin{eqnarray}
\langle \hat{V} \rangle_{\gamma}&=&\alpha(\gamma)\langle \hat{V} \rangle_{\gamma=1} \nonumber
\end{eqnarray}
is valid for the Coulomb energy of any state which is rotationally invariant about the origin in the scaled coordinates, and does not require either homogenous density or translational invariance. (Note: In this section all coordinates stand for the scaled coordinates, but we drop the subscript $s$ for ease of notation.) Examples of such a state include a CF hole or a CF particle located at the origin, or even the ground state for a finite system. We use the following notations to denote the coordinates: $\vec{r}_{1}$ and $\vec{r}_{2}$ are used to denote the coordinates $(x_1, y_1)$ and $(x_2, y_2)$ of a pair of particles in the scaled frame. The center of mass and relative coordinates are defined as $\vec{R}=(\vec{r}_1+\vec{r}_2)/2\equiv (R,\delta)$ and $\vec{r}=\vec{r}_1-\vec{r}_2\equiv(r,\theta)$, respectively. We also use complex coordinates in this section defined as $z_1=x_1-iy_1$, $z_2=x_2-iy_2$, $z=z_1-z_2$, and $Z=(z_1+z_2)/2$.

Since the state is rotationally invariant about the origin (in the scaled frame), the densities $\rho(\vec{r})$ and $\rho_b(\vec{r})$ only depend on the absolute value of $\vec{r}$, i.e., $\rho(\vec{r})=\rho(r)$ and $\rho_b(\vec{r})=\rho_b(r)$. Let us first look at the background-background interaction:
\begin{eqnarray}
\langle \hat{V}_{\rm bb} \rangle&=&{e^2\over 2\epsilon} \int d\vec{R} \int d\vec{r}~\frac{\rho_{b}(|2\vec{R}+\vec{r}|/2)\rho_{b}(|2\vec{R}-\vec{r}|/2)}{\sqrt{(\sqrt{\gamma}r\cos\theta)^2+(\frac{r\sin\theta}{\sqrt{\gamma}})^2}} \nonumber 
\end{eqnarray}
where we have used the fact that the quantities $r_{1}$ and $r_{2}$ are given by:
\begin{eqnarray}
2r_1=|2\vec{R}+\vec{r}|&=&\sqrt{4R^2+r^2+4Rr\cos(\delta-\theta)} \nonumber \\
2r_2=|2\vec{R}-\vec{r}|&=&\sqrt{4R^2+r^2-4Rr\cos(\delta-\theta)} 
\label{r_1_and_r_2}
\end{eqnarray}
Defining $\eta_{\rm bb}(\vec{R},\vec{r})=\rho_{b}(r_{1})\rho_{b}(r_{2})\equiv \eta_{\rm bb}(R,r,\cos(\delta-\theta))$ we have
\begin{eqnarray}
\langle \hat{V}_{\rm bb} \rangle&=&{e^2\over 2\epsilon} \int d\vec{R}~\int d\vec{r}~\frac{\eta_{\rm bb}(R,r,\cos(\delta-\theta))}{\sqrt{(\sqrt{\gamma}r\cos\theta)^2+(\frac{r\sin\theta}{\sqrt{\gamma}})^2}}
\end{eqnarray}
Substituting $\delta-\theta=\beta$ gives:
\begin{eqnarray}
\langle \hat{V}_{\rm bb} \rangle&=&{e^2\over 2\epsilon} \int_{0}^{\infty} R dR~\int_{0}^{\infty} r dr~\int_{0}^{2\pi}d\theta~\frac{1}{\sqrt{(\sqrt{\gamma}r\cos\theta)^2+(\frac{r\sin\theta}{\sqrt{\gamma}})^2}}~\Bigg\{\int_{-\theta}^{2\pi-\theta}d\beta~\eta_{\rm bb}(R,r,\cos(\beta))\Bigg\} \nonumber \\
\end{eqnarray}
Since $\eta_{\rm bb}$ only depends on $\cos\beta$ it is a periodic function of $\beta$ with period $2\pi$. We then make the reasonable assumption that the background density profile is such that $\eta_{\rm bb}$ is a continuous function of $\beta$ which allows us to change the limits of the integral to zero to $2\pi$. The term in the curly parentheses thus depends only on $R$ and $r$ and not on $\theta$. Calling this term $h(R,r)$, we get
\begin{eqnarray}
\langle \hat{V}_{\rm bb} \rangle&=&{e^2\over 2\epsilon} \int_{0}^{\infty} R dR~\int_{0}^{\infty} dr~h(R,r)~\int_{0}^{2\pi}d\theta~\frac{1}{\sqrt{(\sqrt{\gamma}\cos\theta)^2+(\frac{\sin\theta}{\sqrt{\gamma}})^2}}=\alpha(\gamma) \langle \hat{V}_{\rm bb} \rangle_{\gamma=1}
\label{bb_rot_inv}
\end{eqnarray}

Exactly in the same fashion, it can be shown that the electron-background term:
\be
\langle \hat{V}_{\rm eb} \rangle=-{e^2} \int d\vec{r}_{1} \int d\vec{r}_{2}~\rho(r_{1})~\rho_{b}(r_{2}) V(\vec{r}_{1},\vec{r}_{2})
\ee
also satisfies 
\begin{eqnarray}
\langle \hat{V}_{\rm eb} \rangle&=\alpha(\gamma) \langle \hat{V}_{\rm eb} \rangle_{\gamma=1}
\label{eb_rot_inv}
\end{eqnarray}

Finally we look at the electron-electron interaction term:
\be
\langle \hat{V}_{\rm ee} \rangle={e^2} \int d\vec{r}_{1} \int d\vec{r}_{2}~\rho(r_{1}) \rho(r_{2}) g(\vec{r}_{1},\vec{r}_{2}) V(\vec{r}_{1},\vec{r}_{2})
\ee
The pair-correlation function is closely related to the off-diagonal density matrix which is defined as:
\begin{eqnarray}
{\langle \hat{\Psi}^{\dagger}(\vec{r}_{1})  \hat{\Psi}^{\dagger}(\vec{r}_{2})  \hat{\Psi}(\vec{r}'_{2}) \hat{\Psi}(\vec{r}'_{1})    \rangle}&=&N(N-1)\int d^{2}\vec{r}_{3}\dots d^2\vec{r}_{N}~\Phi^{*}(z_{1},z_{2},z_{3},\dots,z_{N})\Phi(z_{1}',z_{2}',z_{3},\dots,z_{N}) \nonumber \\
 &\equiv& f_{1}(z_{1}^{*},z_{2}^{*})~f_{2}(z_{1}',z_{2}') \exp\left(- \frac{|z_{1}|^2}{4l^2}-\frac{|z_{2}|^2}{4l^2}-\frac{|z_{1}'|^2}{4l^2}-\frac{|z_{2}'|^2}{4l^2} \right)
\label{eq1}
\end{eqnarray}
where $\Phi$ is the $N$ electron wave function. The last step follows from the analytic form of LLL wave functions. The functional forms of $f_{1}$ and $f_{2}$ are the same. Noting that the functions $f_{1}$ and $f_{2}$ are antisymmetric functions of their arguments, we can write the power series expansion:
\begin{eqnarray}
{\langle \hat{\Psi}^{\dagger}(\vec{r}_{1})  \hat{\Psi}^{\dagger}(\vec{r}_{2})  \hat{\Psi}(\vec{r}'_{2}) \hat{\Psi}(\vec{r}'_{1})    \rangle}&=&\sum_{m_{1},n_{1},m_{2},n_{2}} c_{m_{1},n_{1},m_{2},n_{2}} (z_{1}^{*}-z_{2}^{*})^{m_1} (z_{1}'-z_{2}')^{m_2}(z_{1}^{*}+z_{2}^{*})^{n_1} (z_{1}'+z_{2}')^{n_2} \nonumber \\
&& \times \exp\left[- \frac{|z_{1}|^2}{4l^2}-\frac{|z_{2}|^2}{4l^2}-\frac{|z_{1}'|^2}{4l^2}-\frac{|z_{2}'|^2}{4l^2} \right] \nonumber
\end{eqnarray}
where $m_{1}$ and $m_{2}$ are odd positive integers and $n_{1}$ and $n_{2}$ are non-negative integers. Now we set $z_{1}=z_{1}'$ and $z_{2}=z_{2}'$ to obtain
\begin{eqnarray}
{\langle \hat{\Psi}^{\dagger}(\vec{r}_{1})  \hat{\Psi}^{\dagger}(\vec{r}_{2})  \hat{\Psi}(\vec{r}_{2}) \hat{\Psi}(\vec{r}_{1})    \rangle}
&=&\sum_{m_{1},n_{1},m_{2},n_{2}} c_{m_{1},n_{1},m_{2},n_{2}} \left[(z^{*})^{m_1} (z)^{m_2}(Z^{*})^{n_1} (Z)^{n_2}+c.c.\right]F(R,r)
 \nonumber \\
\end{eqnarray}
where we have made use of the reality of the expectation value and defined $F(R,r)=\exp[-(4R^2+r^2)/(4l^2)]$.

So far the discussion is general for any wave function in the lowest LL. Now we ask what simplification is produced by global rotational symmetry of the wave functions, which implies that the pair correlation function be invariant under the transformation: $z\rightarrow ze^{i\psi}$, $Z\rightarrow Ze^{i\psi}$ in the scaled coordinates. This implies that 
\begin{equation}
m_{1}+n_{1}=m_{2}+n_{2} 
\label{condition1}
\end{equation}
Defining $\eta(\vec{r},\vec{R})=\rho(r_{1})\rho(r_{2})\equiv \eta(R,r,\cos(\delta-\theta))$ we find:
\begin{eqnarray}
\langle \hat{V}_{\rm ee} \rangle&=&{e^2\over 2} \int d\vec{R}~d\vec{r}~V(\vec{r}) \eta(\vec{r},\vec{R})g(\vec{r},\vec{R}) \nonumber \\
&=&{e^2\over 2} \sum_{m_{1},m_{2},n_{1},n_{2}} c_{m_{1},m_{2},n_{1},n_{2}} \int d\vec{r}~V(\vec{r})~\int_{0}^{\infty}RdR~\int_{0}^{2\pi}d\delta  \nonumber \\
&& \times~r^{m_{1}+m_{2}}R^{n_{1}+n_{2}}
~2\cos[(m_{2}-m_{1})\theta + (n_{2}-n_{1})\delta]\eta(r,R,\cos(\delta-\theta))F(R,r) \nonumber
\end{eqnarray}
With the substitution $\beta=\delta-\theta$ we get
\begin{eqnarray}
\langle \hat{V}_{\rm ee} \rangle&={e^2\over 2} &\sum_{m_{1},m_{2},n_{1},n_{2}} c_{m_{1},m_{2},n_{1},n_{2}} \int d\vec{r}~V(\vec{r})~\int_{0}^{\infty}RdR~F(R,r)r^{m_{1}+m_{2}}R^{n_{1}+n_{2}} \nonumber \\
&&
\Bigg\{\times \int_{-\theta}^{2\pi-\theta}d\beta 
~2\cos[(m_{2}-m_{1}+n_{2}-n_{1})\theta + (n_{2}-n_{1})\beta] 
\eta(r,R,\cos(\beta))\Bigg\} \nonumber
\end{eqnarray}
Using the constraint given in Eq. \ref{condition1} and making the reasonable assumption that $\eta$ is a continuous function of $\beta$ we change the limits of the integral inside the curly brackets as before to zero to $2\pi$. The expression on the right-hand side then takes the form
\begin{eqnarray}
\langle \hat{V}_{\rm ee} \rangle&=&{e^2 \over 2}\sum_{m_{1},m_{2},n_{1},n_{2}} c_{m_{1},m_{2},n_{1},n_{2}}~\int_{0}^{\infty}RdR~\int d\vec{r}~V(\vec{r})H(R,r) 
\end{eqnarray}
which implies 
\begin{eqnarray}
\langle \hat{V}_{\rm ee} \rangle&=& \alpha(\gamma)\langle \hat{V}_{\rm ee} \rangle_{\gamma=1}
\label{ee_rot_inv}
\end{eqnarray}
Combining Eqs. \ref{bb_rot_inv}, \ref{eb_rot_inv} and \ref{ee_rot_inv} gives Eq.~\ref{tot_rot_inv}.
We reiterate the fact that this result is true for a finite system, and only assumes rotational invariance in the scaled reference frame. 

\end{widetext}
\end{appendices}

\bibliography{../../Latex-Revtex-etc./biblio_fqhe}
\bibliographystyle{apsrev}

\end{document}